\documentclass[article,ij4uq]{ij4uq}              

 \usepackage{amsmath}
\usepackage[hang]{footmisc}
\fancypagestyle{plain}{%
  \fancyhf{}
  \fancyhead[R]{\small {\it International Journal for Uncertainty Quantification}, x(x): \thepage--\pageref{LastPage} (\myyear\today)}
  \fancyfoot[R]{\small\bf\thepage }
  \fancyfoot[L]{\fottitle}
  }
\renewcommand{\dmyy}{17}
\renewcommand{\myyear}{2017}
\renewcommand{\today}{}
\begin{document}

\volume{Volume x, Issue x, \myyear\today}
\title{Adaptive selection of sampling points for Uncertainty Quantification} 
\titlehead{Adaptive sampling}
\authorhead{E. Camporeale, A.  Agnihotri, \& C. Rutjes}
\corrauthor[1]{Enrico Camporeale}
\author[1]{Ashuthosh Agnihotri}
\author[1]{Casper Rutjes}
\corremail{e.camporeale@cwi.nl}
\corraddress{Center for Mathematics and Computer Science (CWI), Amsterdam, The Netherlands}

\dataO{mm/dd/yyyy}
\dataF{mm/dd/yyyy}

\abstract{We present a simple and robust strategy for the selection of sampling 
points in Uncertainty Quantification. The goal is to achieve the fastest possible convergence in the cumulative distribution function of a stochastic output of interest.
We assume that the output of interest is the outcome of a computationally expensive nonlinear mapping of an input random variable, whose probability density function is known.
We use a radial function basis to construct an accurate interpolant of the mapping.
This strategy  enables adding new sampling points one at a time, \emph{adaptively}. This takes into full account the previous evaluations of the target nonlinear function. 
We present comparisons with a stochastic collocation method based on the Clenshaw-Curtis quadrature rule, and with an adaptive method based on hierarchical surplus, showing that the new method often results in a large computational saving.

}


\maketitle

\section{Introduction}
We address one of the fundamental problems in Uncertainty Quantification (UQ): the mapping of the probability distribution of a random variable through a nonlinear function.
Let us assume that we are concerned with a specific physical or engineering model which is computationally expensive. The model is defined by the map $ g : \mathbb{R} \rightarrow \mathbb{R} $. 
It takes a parameter $ X $ as input, and produces an output $ Y $, $ Y = g(X) $. In this paper we restrict ourselves to a proof-of-principle one-dimensional case. 
Let us assume that $ X $ is a random variable distributed with probability density function (pdf) $ P_{X} $. 
The Uncertainty Quantification problem is the estimation of the pdf $ P_{Y} $ of the output variable $Y$, given $ P_{X} $.
Formally, the problem can be simply cast as a coordinate transformation and one easily obtains
\begin{equation}
 P_Y(y) = \sum_{x \in \{ x | g(x) = y\}}\frac{P_X(x)}{|\det J(x)|},\label{Py}
\end{equation}
where $J(x)$ is the Jacobian of $g({x})$. The sum over all $x$ 
such that $g(x)=y$ takes in account the possibility that $g$ may not be injective. If the function $g$ is known exactly and invertible, Eq.(\ref{Py}) can be used straightforwardly to construct the pdf $P_Y(y)$, but this is of course not the case when the mapping $g$ is computed via numerical simulations.

Several techniques have been studied in the last couple of decades to tackle this problem. Generally, the techniques can be divided in two categories: intrusive and non-intrusive \cite{xiu09,eldred09,onorato2010comparison}. Intrusive methods modify the original, 
{\it deterministic}, set of equations to account for the stochastic nature of the input (random) variables, hence eventually dealing with stochastic differential equations, and employing specific numerical techniques  to solve them. Classical examples of intrusive methods are represented by Polynomial Chaos expansion \cite{xiu02,crestaux09,Togawa:2011,eldred2009recent}, and stochastic Galerkin methods \cite{link2,grigoriu2012stochastic,xiu2010numerical,le10}.

On the other hand, the philosophy behind non-intrusive methods is to make use of the deterministic version of the model (and the computer code that solves it) as a black-box, which returns one deterministic output for any given input. An arbitrary large number of solutions, obtained by sampling the input parameter space, can then be collected and analyzed in order to reconstruct the pdf $P_Y(y)$.

The paradigm of non-intrusive methods is perhaps best represented by 
Monte Carlo (MC) methods \cite{link4,kalos2008monte}: one can construct an ensemble of input parameters $ \{ X_{n} \, | \, n = 1,\dots,N \} $ ($N$ typically large) distributed according to the pdf $P_X(x)$, run the corresponding ensemble of simulations $ g : X \rightarrow Y$, and process the outputs $\{Y_{n} \, | \, n=1,\dots,N\}$. MC methods are probably the most robust of all the non-intrusive methods. Their main shortcoming is the slow convergence of the method, with a typical convergence rate proportional to $\sqrt{N}$. For many applications quasi-Monte Carlo (QMC) methods \cite{link4,l2009monte} are now preferred to MC methods, for their faster convergence rate. In QMC the 
pseudo-random generator of samples is replaced by more uniform distributions, obtained through so-called quasi-random generators \cite{link5,kalos2008pseudorandom}. 

It is often said that MC and QMC do not suffer the `curse of dimensionality'\cite{link6,indyk1998approximate,kuo2005lifting}, in the sense that the convergence rate (but not the actual error!) is not affected by the dimension $D$ of the input parameter space. Therefore, they represent the standard choice for large dimensional problems. On the other hand, when the dimension $D$ is not very large, collocation methods \cite{link7,Foo2010,babuvska2007stochastic} are usually more efficient.

Yet a different method that focuses on deriving a deterministic differential equation for cumulative distribution functions has been presented, e.g., in \cite{wang12, wang13}. This method is however not completely black-box.

Collocation methods recast an UQ problem as an interpolation problem.
In collocation methods, the function $g(x)$ is sampled in a small 
(compared to the MC approach) number of points (`collocation points'),
 and an interpolant is constructed to obtain an approximation of $g$ over 
 the whole input parameter space, from which the pdf $P_Y(y)$ can be estimated. 
 
The question then arises on how to effectively choose the collocation points.
Recalling that every evaluation of the function $g$ amounts to performing an expensive simulation, the challenge resides in obtaining an accurate approximation of $P_Y$ with the least number of collocation points. Indeed, a very active area of research is represented by collocation methods that use sparse grids, so to avoid the computation of a full-rank tensorial product, particularly for model order reduction (see, e.g., \cite{smolyak63, ganapathysubramanian07, bieri09, ma09b, link17, jakeman12, nguyen12} 
\\ 
As the name suggests, collocation methods are usually derived from classical quadrature rules \cite{link8,gautschi1970construction,waldvogel2006fast}. 

The type of pdf $P_X$ can guide the choice of the optimal quadrature rule to be used (i.e., Gauss-Hermite for a Gaussian probability, Gauss-Legendre for a uniform probability, etc. \cite{link7}). Furthermore, because quadratures are associated with polynomial interpolation, it becomes natural to define a global interpolant in terms of a Lagrange polynomial \cite{link9}. Also, choosing the collocation points as the abscissas of a given quadrature rule makes sense particularly if one is only interested in the evaluation of the statistical moments of the pdf (i.e., mean, variance, etc.) \cite{link19}.

On the other hand, there are several applications where one is interested in the approximation of the full pdf $P_Y$. For instance, when $g$ is narrowly peaked around two or more distinct values, its mean does not have any statistical meaning. In such cases one can wonder whether a standard collocation method based on quadrature rules still represents the optimal choice, in the sense of the computational cost to obtain a given 
accuracy. 

From this perspective, a downside of collocation methods is that the collocation points are chosen a priori, without making use of the knowledge of $g(x)$ acquired at previous interpolation levels. 
For instance, the Clenshaw-Curtis (CC) method uses a set of points that contains 'nested' subset, in order to re-use all the previous computations, when the number of collocation points is increased.
However, since the abscissas are unevenly spaced and concentrated towards the edge of the domain (this is typical of all quadrature rules, in order to overcome the Runge 
phenomenon \cite{link9,link11}), it is likely that the majority of the 
performed simulations will not contribute significantly in achieving a better approximation of $P_Y$. Stated differently, one would like to employ a method where each new sampling point is chosen in such a way to result in the fastest convergence rate for the approximated $P_Y$, in contrast to a set of points defined a priori. 

As a matter of fact, because the function $g$ is unknown, a certain number of simulations will always be redundant, in the sense that they will contribute very little to the convergence of $ P_{Y} $. The rationale for this work is to devise a method to minimize such a redundancy in the choice of sampling points while achieving fastest possible convergence of $ P_{Y} $. 

Clearly, this suggests to devise a strategy that chooses collocation points {\it adaptively}, making use of the knowledge of the interpolant of $g(x)$, which becomes more and more accurate as more points are added.

\begin{figure}
\includegraphics[width=10 cm]{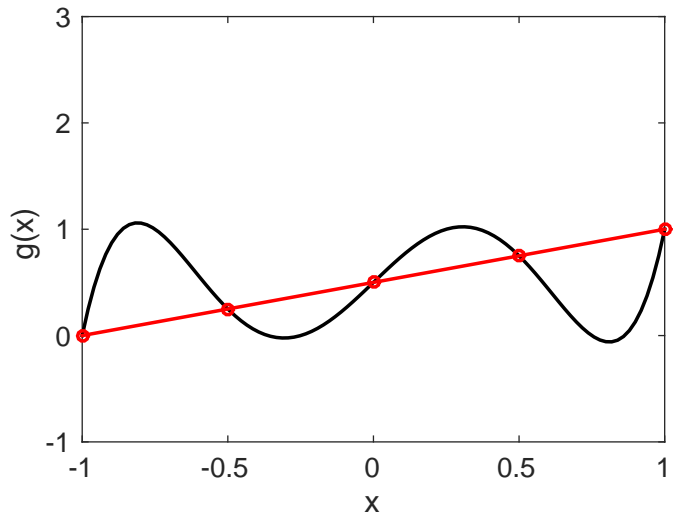}
\caption{Example for which the algorithm based on hierarchical surplus fails. The function $g(x)= \frac{256}{30}x^5 - \frac{32}{3}x^3 +\frac{79}{30}x + \frac{1}{2}$ (in black) goes exactly through the red straight line at the points $x=-1,-0.5,0,0.5,1$. Calculating the piece-wise linear interpolant between two ($x=-1,1$), three ($x=-1,0,1$), and five ($x=-1,-\frac{1}{2},0,\frac{1}{2},1$) points would result in a null hierarchical surplus on these points.} \label{fig:hierarchical_example}
\end{figure}

A well known adaptive sampling algorithm is based on the calculation of the so-called hierarchical surplus \cite[see e.g]{ma09,ma09b,jakeman12,witteveen12}. 
This is defined as the difference, between two levels of refinement, in the solution obtained by the interpolant. Although this algorithm is quite robust, and it is especially efficient 
in detecting discontinuities, it has the obvious drawback that it can be prematurely terminated, whenever the interpolant happens to exactly pass through the true solution on a point where the hierarchical surplus is calculated, no matter how inaccurate the interpolant is in close-by regions (see Figure \ref{fig:hierarchical_example} for an example).

The goal of this paper is to describe an alternative strategy for the adaptive selection of sampling points. The objective in devising such strategy is to have a simple and robust set of rules for choosing the next sampling point. The paper is concerned with a proof-of-principle demonstration of our new strategy, and we will focus here on one dimensional cases and on the case of uniform $ P_{X} $ only, postponing the generalization to multiple dimensions to future work. 
It is important to appreciate that the stated goal of this work is different from the traditional approach followed in the overwhelming majority of works that have presented sampling methods for UQ in the literature. Indeed, it is standard to focus on the convergence of the nonlinear unknown function $g(x)$, trying to minimize the interpolation error on $g(x)$, for a given number of sampling points. On the other hand, we will show that the convergence rates of $g(x)$ and of its cumulative distribution function can be quite different. Our new strategy is designed to achieve the fastest convergence on the latter quantity, which is ultimately the observable quantity of an experiment.

The paper is organized as follows. In Section 2 we define the mathematical methods used for the construction of the interpolant and show our adaptive strategy to choose a new collocation points. In Section 3 we present some numerical examples and comparisons with the Clenshaw-Curtis collocation method, and the adaptive method based on hierarchical surplus. 
Finally, we draw our conclusions in Section 4.

\section{Mathematical methods}\label{sec:math}

\subsection{Clenshaw-Curtis (CC) quadrature rule}
In Section 3, we compare our method with the CC method, which is the standard appropriate collocation method for a uniform $P_X$.
Here, we recall the basic properties of CC, for completeness.
The Clenshaw-Curtis (CC) quadrature rule uses the extrema of a Chebyshev 
polynomial (the so-called `extrema plus end-points' collocation points in 
\cite{link10}) as abscissas. They are particularly appealing to be used as 
collocation points in UQ, because a certain subset of them are 
nested. Specifically, they are defined, in the interval $[-1,1]$ as:
\begin{equation}
 x_i = -\cos\left(\frac{\pi (i-1)}{N-1}\right) \quad i=1,\dots,N.
\end{equation}
One can notice that the the set of $N=2^w+1$ points is fully contained in the 
set of $N=2^{w+1}+1$ points (with $w$ an arbitrary integer, referred to as the 
level of the set). In practice this means that one can construct a nested 
sequence of collocation points with $N=3,5,9,17,33,65,129,\dots$, re-using all 
the previous evaluations of $g$. 

Collocation points based on quadratures are optimal to calculate moments \footnote{Here $ p$ on the left-hand side is a label, such that $ \mu^{1}  $ is the mean, $ \mu^{2} $ is the variance, and so on. On the right-hand side it is an exponent.}: 
\begin{equation}\label{eq:mean}
 \mu_Y^p=\int y^p P_Y(y) {\rm d}y = \int g(x)^p P_X(x) {\rm d}x,
\end{equation}
where we used the identity relation,
\begin{equation}
  P_Y(y) {\rm d}y =  P_X(x) {\rm d}x.
\end{equation}
It is known that integration by quadrature is very accurate (for 
smooth enough integrand), and the moments can be readily evaluated, without the 
need to construct an interpolant:
\begin{equation}
 \mu_Y^p\simeq \sum_i w_i (g(x_i))^p,
\end{equation}
where the weights $w_i$ can be computed with standard techniques (see, e.g. 
\cite{link19}).
The interpolant for the CC method is the Lagrange polynomial.

\subsection{Selection of collocation points based on hierarchical surplus}
The hierarchical surplus algorithm is widely used for interpolation on sparse grids. It is generally defined as the difference
between the value of an interpolant at the current and previous interpolation levels \cite{ma09}:
\begin{equation}
 \Delta^n = \tilde{g}^{n}-\tilde{g}^{n-1}
\end{equation}

The simplest algorithm prescribes a certain tolerance
and looks for all the point at the new level where the hierarchical surplus is larger than the tolerance. The new sampling points (at the next level, $n+1$) will be the neighbors (defined with a certain rule)
of the points where this condition is met.
In one-dimension, the algorithm is extremely simple because the neighbors are defined by only two points, that one can define in such a way that cells are always halved.
In this work, we compare our new method with a slightly improved version of the hierarchical surplus algorithm. The reason is because we do not want our comparisons to be dependent on 
the choice of an arbitrary tolerance level, and we want to be able to add new points two at the time.
Hence, we define a new interpolation level by adding only the two neighbors of the point with the largest hierarchical surplus. All the previous hierarchical surpluses that have been calculated, but for which new points have not been added yet are kept.
The pseudo-code of the algorithm follows. The interpolant is understood to be piece-wise linear interpolation, and the grid is $x\in[-1,1]$.
\begin{algorithm}
\caption{Hierarchical surplus algorithm}
Calculate the interpolant on the grid $x=\{-1,0,1\}$.\\
Define $x_{h}=\{-1/2, 1/2\}$ and add them on the grid\\
\While {Not converged}{ 
    Calculate the interpolant on the new grid\\
    Calculate the hierarchical surplus on the last two entries of $x_{h}$ and store them in the vector $h_s$ \\
    Find the largest hierarchical surplus in $h_s$, remove it from $h_s$ and remove the corresponding $x$ from $x_h$ \\
    Append the two neighbors to $x_h$ and add them to the grid
}
\end{algorithm}

\subsection{Multiquadric biharmonic radial basis}
We use a  multiquadric biharmonic radial basis function (RBF) with respect to a 
set of points $\{x_i\}$, with $i=1,\dots,N$, defined as:
\begin{equation}
 \Phi_i(x,c)=\sqrt{(x-x_i)^2+c_i^2},
\end{equation}
where $c_i$ are free parameters (referred to as shape parameters).
The function $g(x)$ is approximated by the interpolant $\widetilde{g}(x)$ 
defined as
\begin{equation}
 \widetilde{g}(x) = \sum_{i=1}^N\lambda_i\Phi_i(x,c).\label{interpolant}
\end{equation}
The weights $\lambda_i$ are obtained by imposing that 
$g(x_i)=\widetilde{g}(x_i)$ for each sampling point in the set, namely the 
interpolation error is null at the sampling points. This results in solving a 
linear system for ${\bf \lambda}=(\lambda_1,\dots,\lambda_N)$ of the form $A{\bf 
\lambda}^T=g({\bf x})^T$, with $A$ a real symmetric $N\times N$ matrix. 
We note that, by construction, the linear system will become more and more 
ill-conditioned with increasing $N$, for fixed values of $c$. This can be easily 
understood because when two points become closer and closer the corresponding 
two rows in the matrix $A$ become less and less linearly independent. To 
overcome this problem one needs to decrease the corresponding values of $c$. In 
turns, this means that the interpolant $\widetilde{g}(x)$ will tend to a 
piece-wise linear interpolant for increasingly large $N$.

\subsection{New adaptive selection of collocation points}
We focus, as the main diagnostic of our method, on the cumulative distribution 
function (cdf) $C(y)$, which is defined as
\begin{equation}
 C(y)=\int_{y_{\rm min}}^y P_Y(y) {\rm d}y,
\end{equation}
where $y_{\rm min} = \min g(x)  $. As it is well known, the interpretation of the 
cumulative distribution function is that, for a given value $ y^* $, $ C(y^*) $ is the probability that $ g(x) \leq y^* $.
Of course, the cdf $C(y)$ contains all the statistical information needed to 
calculate any moment of the distribution, and can return the probability density 
function $P_Y(y)$, upon differentiation. Moreover, the cdf is always well 
defined between 0 and 1.
The following two straightforward considerations will guide the design of our 
{\it adaptive selection strategy}. 
A first crucial point, already evident from Eq. (\ref{Py}), is whether or not 
$g(x)$ is bijective. 
When $g(x)$ is bijective this translates to the cdf $C(y)$ being continuous, while a 
non-bijective function $g(x)$ produces a cdf $C(y)$ which is discontinuous. 
It follows that intervals in $x$ where $g(x)$ is constant (or nearly constant) 
will map into a single value $y=g(x)$ (or a very small interval in $y$) where 
the cdf will be discontinuous (or `nearly' discontinuous). Secondly, an interval 
in $x$ with a large first derivative of $g(x)$ will produce a nearly flat 
cdf $C(y)$. This is again clear by noticing that the Jacobian $J$ in Eq. (\ref{Py}) 
(${\rm d} g(x)/ {\rm d}x$ in one dimension) is in the denominator, and therefore the 
corresponding $P_Y(y)$ will be very small, resulting in a flat cdf $C(y)$.\\
Loosely speaking one can then state that regions where $g(x)$ is flat will 
produce large jumps in the cdf $C(y)$ and, conversely, regions where the $g(x)$ has 
large jumps will map in to a nearly flat cdf $C(y)$. From this simple considerations 
one can appreciate how important it is to have an interpolant that accurately 
capture both regions with {\it very large} and {\it very small} first derivative 
of $g(x)$. Moreover, since the cdf $C(y)$ is an integrated quantity, interpolation 
errors committed around a given $y$ will propagate in the cdf for all larger 
$y$ values. For this reason, it is important to achieve a global convergence 
with interpolation errors that are of the same order of magnitude along the 
whole domain.\\
The adaptive section algorithm works as follows. We work in the interval 
$x\in[-1,1]$ (every other interval where the support of $g(x)$ is defined can be 
rescaled to this interval). We denote with $\{x_i\}$ the sampling set which we 
assume is always sorted, such that $x_{i}<x_{i+1}$. We start with 3 points: 
$x_1=-1$, $x_2=0$, $x_3=1$.
For the robustness and the simplicity of the implementation we choose to select 
a new sampling point always at equal distance between two existing points. One 
can decide to limit the ratio between the largest and smallest distance between 
adjacent points: $r=\max\{d_i\}/\min\{d_i\}$  (with $i=1,\dots,N-1$), where 
$d_i$ is the distance between the points $x_{i+1}$ and $x_i$. This avoids to 
keep refining small intervals when large intervals might still be 
under-resolved, thus aiming for the above mentioned global convergence over the 
whole support.
At each iteration we create a list of possible new points, by halving every 
interval, excluding the points that would increase the value of $r$ above the 
maximum desired (note that $r$ will always be a power of 2). We calculate the 
first derivative of $\widetilde{g}(x)$ at these points, and alternatively choose 
the point with largest/smallest derivative as the next sampling point. Notice 
that, by the definition of the interpolant, Eq. (\ref{interpolant}), its first 
derivative can be calculated exactly as:
\begin{equation}\label{first_derivative}
\frac{{\rm d}\widetilde{g}(x)}{{\rm d}x}=\sum_{i=1}^N\lambda_i\frac{{\rm d}\Phi_i(x,c)}{{\rm d}x}
\end{equation}
without having to recompute the weights $\lambda_i$. At each iteration the shape 
parameters $c_i$ are defined at each points, as 
$c_i=0.85\cdot\min(d_{i-1},d_{i})$, i.e. they are linearly rescaled with the 
smallest distance between the point $x_i$ and its neighbors.
The pseudo-code of the algorithm follows.
\begin{algorithm}
\caption{Adaptive selection of sampling points}
\While {Not converged}{ 
    $x_{\rm guess}\gets 0.5\cdot(x_i+x_{i+1})$\\
    Exclude points in $x_{\rm guess}$ such that $r=\max\{d_i\}/\min\{d_i\} > R$ \\ 
    Calculate $\widetilde{g}^n(x)'$ through (\ref{first_derivative}) at $\{x_{\rm guess}\}$\\
    Alternatively choose $x_{\rm guess}$ with largest/smallest values of $|\widetilde{g}^n(x)'|$ as new collocation point\\
    Calculate new weights $\lambda_i$
}
\end{algorithm}

\section{Numerical examples}
In this section we present and discuss four numerical examples where we apply 
our adaptive selection strategy. In this work we focus on a single input parameter
and the case of constant probability 
$P_X=1/2$ in the interval $x \in [-1,1]$, and we compare our results against the 
Clenshaw-Curtis, and the hierarchical surplus methods. We denote with $\widetilde{g}^n(x)$ the interpolant 
obtained with a set of $n$ points (hence the iterative procedure starts with 
$\widetilde{g}^3(x)$). A possible way to construct the cdf $C(y)$ from a given 
interpolant $\widetilde{g}^n(x)$ would be to generate a sample of points in the 
domain $[-1,1]$, randomly distributed according to the pdf $P_X(x)$, collecting the 
corresponding values calculated through Eq. (\ref{interpolant}), and 
constructing their cdf. Because here we work with a 
constant $P_X(x)$, it is more efficient to simply define a uniform grid in the domain
$ [-1,1] $ where to compute $\widetilde{g}^n(x)$. 
In the following we will use, in the evaluation of the cdf $C(y)$, a grid in $ y $ with 
$N_{y}=10001$ points equally spaced in the interval $[\min 
\widetilde{g}^n(x), \max 
\widetilde{g}^n(x)]$, and a grid in $ x $ with $N_{x}=1001$ points equally 
spaced in the 
interval $[-1,1]$. We define the following errors:
\begin{eqnarray}
 \varepsilon_{C} = \frac{||C(\widetilde{g}^n(x))- 
C(g(x)) ||_2}{\sqrt{N_{y}}}\\
 \varepsilon_{g} = \frac{||\widetilde{g}^n(x) 
-g(x)||_2}{\sqrt{N_{x}}}
\end{eqnarray}
where $||\cdot||_2$ denotes the L$_2$ norm. It is important 
to realize that the accuracy of the numerically evaluated cdf 
$C(y)$ will always 
depend on the binning of $ y$, i.e. the points at which the cdf is 
evaluated. As we will see in the following examples, the error 
$\varepsilon_{C}$ saturates for large $ N $, which thus is an 
artifact of the finite bin size.
We emphasize that, differently from most of the previous literature, our strategy focuses on converging rapidly in $\varepsilon_C$, rather than in $\varepsilon_g$.
Of course, a more accurate interpolant 
will always result in a more accurate cdf, however the relationship between a 
reduction in $\varepsilon_{g}$ and a corresponding reduction in 
$\varepsilon_{C}$ is not at all trivial. This is because the relation between 
$P_X(x)$ and $P_Y(y)$ is mediated by the Jacobian of $g(x)$, and it also 
involves the bijectivity of $g$. \\
Finally, we study the convergence of the 
mean $\mu_Y$, see equation \ref{eq:mean}, and the variance $\sigma_Y^2$, which is defined as
\begin{equation}
 \sigma_{Y}^2 =  \int_{-1}^1 (\widetilde{g}(x)-\mu_Y)^2 P_X(x) 
\textrm{d} x.
\end{equation}
These will be calculated by quadrature for the CC methods, and with an 
integration via trapezoidal method for the adaptive methods.\\
We study two analytical test cases:
\begin{itemize}
 \item Case 1: $g(x) = \arctan(10^3x^3)$;
 \item Case 2: $g(x) = \frac{1}{(2+\sin(3\pi x))^2} $;
\end{itemize}
and two test cases where an analytical solution is not available, and the 
reference $g(x)$ will be calculated as an accurate numerical solution of a set 
of ordinary differential equations:
\begin{itemize}
 \item Case 3: Lotka-Volterra model (predator-prey);
 \item Case 4: Van der Pol oscillator.
\end{itemize}
While Case 1 and 2 are more favorable to the CC method, because the functions 
are smooth and analytical, hence a polynomial interpolation is expected to 
produce accurate results, the latter two cases mimic applications of real 
interest, where the model does not produce analytical results, although $g(x)$ 
might still be smooth (at least piece-wise, in Case 4).

\subsection{Case 1: $g(x) = \arctan(10^3x^3)$}
\begin{figure}
\includegraphics[width=10 cm]{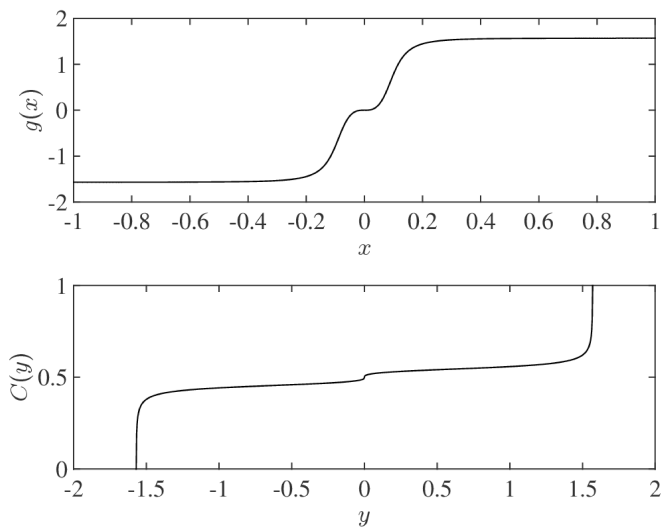}
\caption{Case 1: $g(x) = \arctan(10^3x^3)$. Top panel: $g(x)$; bottom panel: 
cdf $C(y)$.} \label{fig:case_1_f}
\end{figure}

\begin{figure}
\includegraphics[width=10 cm]{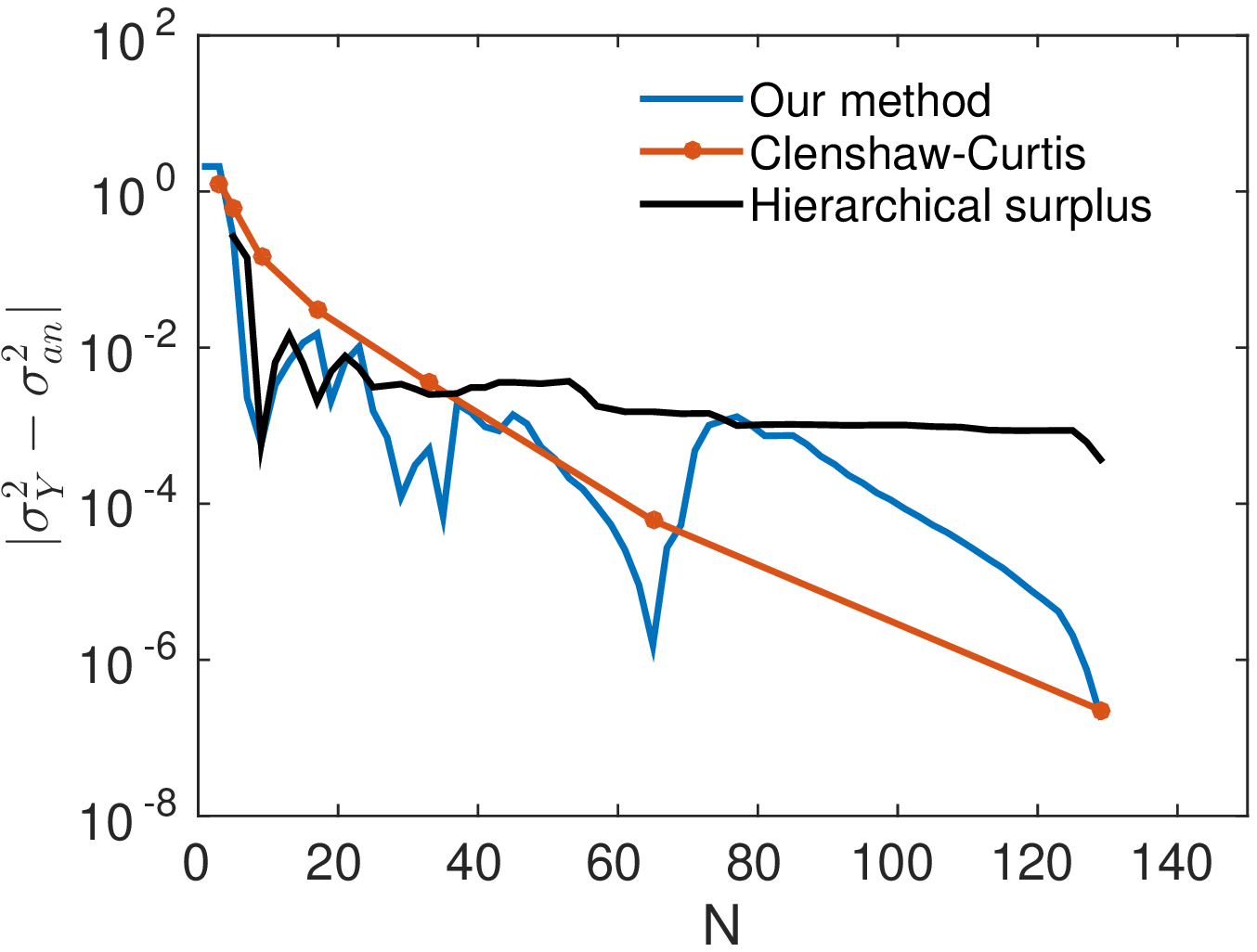}
\caption{Case 1. Absolute error in the variance $\sigma_{Y}^2$ versus number of sampling 
points $N$.  Blue: new adaptive selection method. Red: Clenshaw-Curtis. Black: adaptive method based on hierarchical surplus.} \label{fig:case_1_sigma}
\end{figure}

In this case $g(x)$ is a bijective function, with one point ($x=0$) where the 
first derivative vanishes. Figure \ref{fig:case_1_f} shows the function $g(x)$ (top 
panel) and the corresponding cdf $C(y)$ (bottom panel), which in this case can be 
derived analytically. Hence, we use the analytical expression of cdf $C(y)$ to 
evaluate the error $\varepsilon_{C}$.
The convergence of $\varepsilon_{C}$ and $\varepsilon_{g}$ is shown in Figure \ref{fig:case_1_err} (top and bottom panels, respectively). 
Here and in all the following figures blue squares denote the new adaptive selection 
method, red dots are for the CC methods, and black line is for the hierarchical surplus method. 
We have run the CC method only 
for $N=3, 5, 9, 17, 33, 65, 129$ (i.e. the points at which the collocation 
points are nested), but for a better graphical visualization the red dots are 
connected with straight lines. One can notice that the error for the new adaptive 
method is consistently smaller than for the CC method. From the top panel, one can appreciate the saving in computer power that can be achieved with our 
new method. Although the difference with CC is not very large 
until $N=17$, at $N=33$ there is an order of magnitude difference between the 
two. It effectively means that in order to achieve the same error 
$\varepsilon_{C}\sim10^{-5}$, the CC method would run at least twice the 
number of simulations. 
The importance of focusing on the convergence of the cdf, rather than on the interpolant, is clear in comparing our method with the hierarchical surplus method.
For instance, for $N=80$, the two methods have a comparable error $\varepsilon_g$, but our method has achieved almost an order of magnitude more accurate solution in $C(y)$.
Effectively, this means that our method has sampled the new points less redundantly.
In this case $g(x)$ is an anti-symmetric function with zero mean. Hence, any 
method that chooses sampling points symmetrically distributed around zero would 
produce the correct first moment $\mu_Y$. We show in figure \ref{fig:case_1_sigma} 
the convergence of $\sigma_{Y}^2$, as the absolute value of the different with the exact value $\sigma_{an}$, in logarithmic scale.  
Blue, red, and black lines represent the 
new adaptive method, the CC, and the hierarchical surplus methods, respectively (where again for the CC, simulations are only 
performed where the red dots are shown). The exact value is 
$\sigma_{an}^2=2.102$. 
As we mentioned, the CC method is optimal to calculate moments, since it uses quadrature.
Although in our method the error does not decrease monotonically, it is comparable with the result for CC.

\subsection{Case 2: $g(x) = \frac{1}{(2+\sin(3\pi x))^2} $}

\begin{figure}
\includegraphics[width=10 cm]{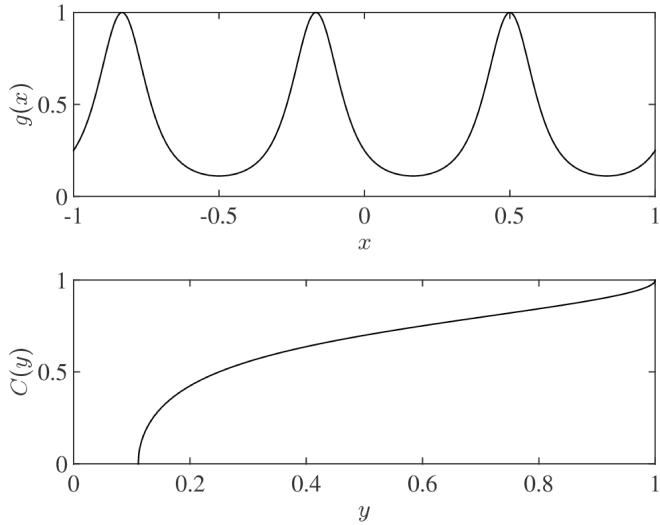}
\caption{Case 2: $g(x) = \frac{1}{(2+\sin(3\pi x))^2} $. Top panel: $g(x)$; 
bottom panel: $C(y)$.} \label{fig:case_2_f}
\end{figure}

\begin{figure}
\includegraphics[width=10 cm]{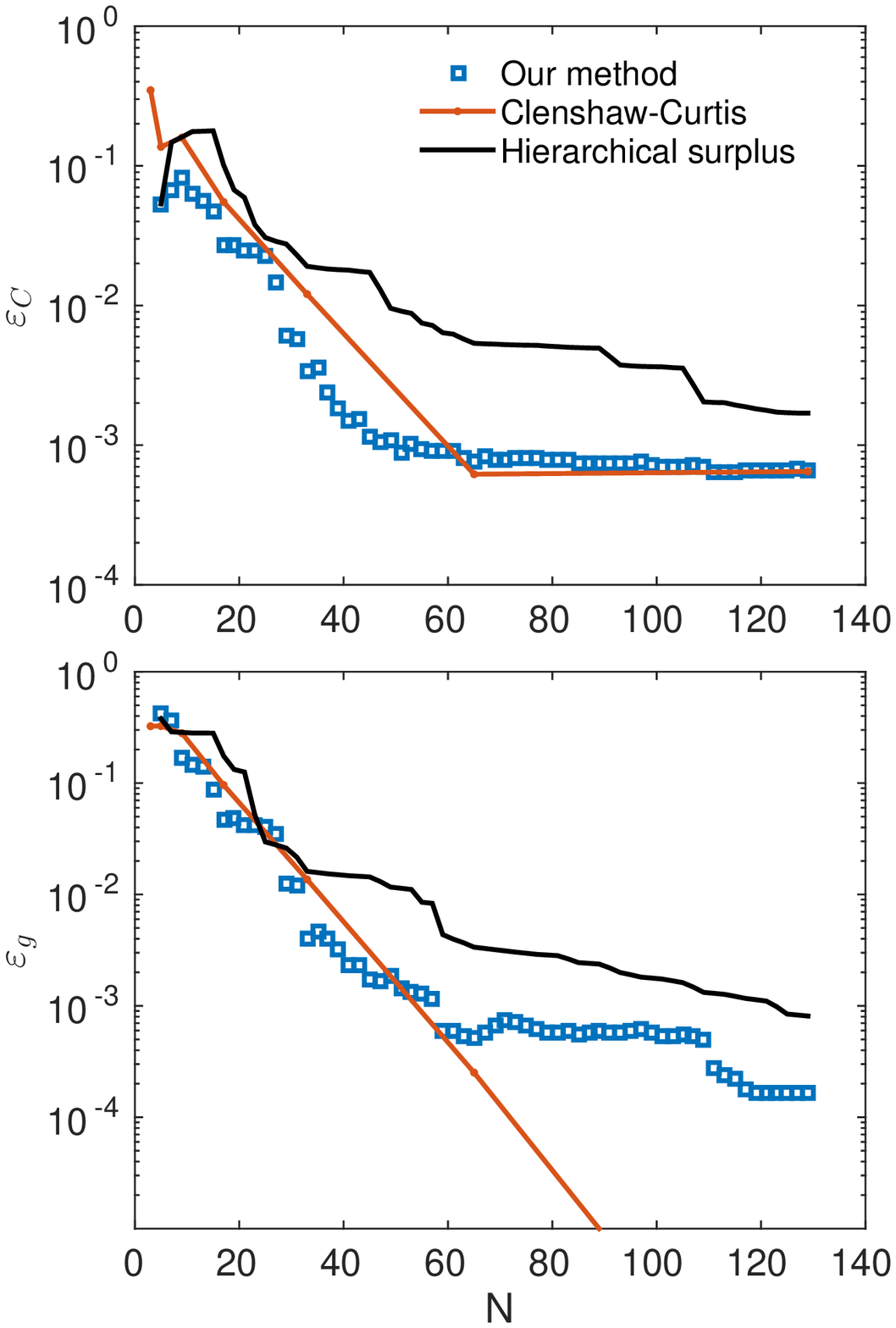}
\caption{Case 2. Error $\varepsilon_{C}$ (top) $\varepsilon_{g}$ (bottom) as function of number of sampling points 
$N$. Blue squares: new adaptive selection method. Red dots: Clenshaw-Curtis. Black curve: adaptive method based on hierarchical surplus.}\label{fig:case_2_err}
\end{figure}

\begin{figure}
\includegraphics[width=10 cm]{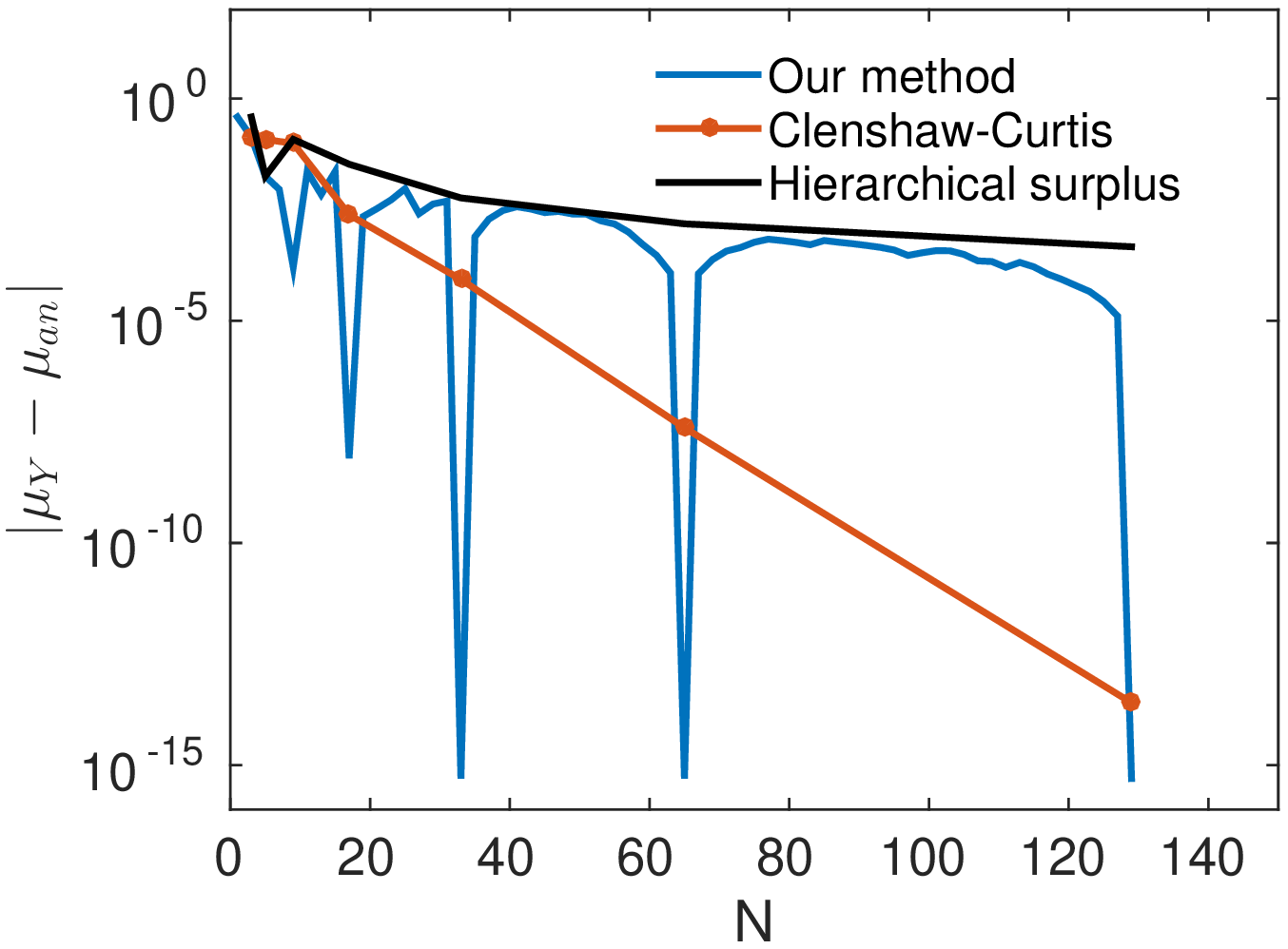}
\caption{Case 2. Absolute error in the mean $\mu_Y$ versus number of sampling 
points $N$.  Blue: new adaptive selection method. Red: Clenshaw-Curtis. Black: adaptive method based on hierarchical surplus.}\label{fig:case_2_mu}
\end{figure}

\begin{figure}
\includegraphics[width=10 cm]{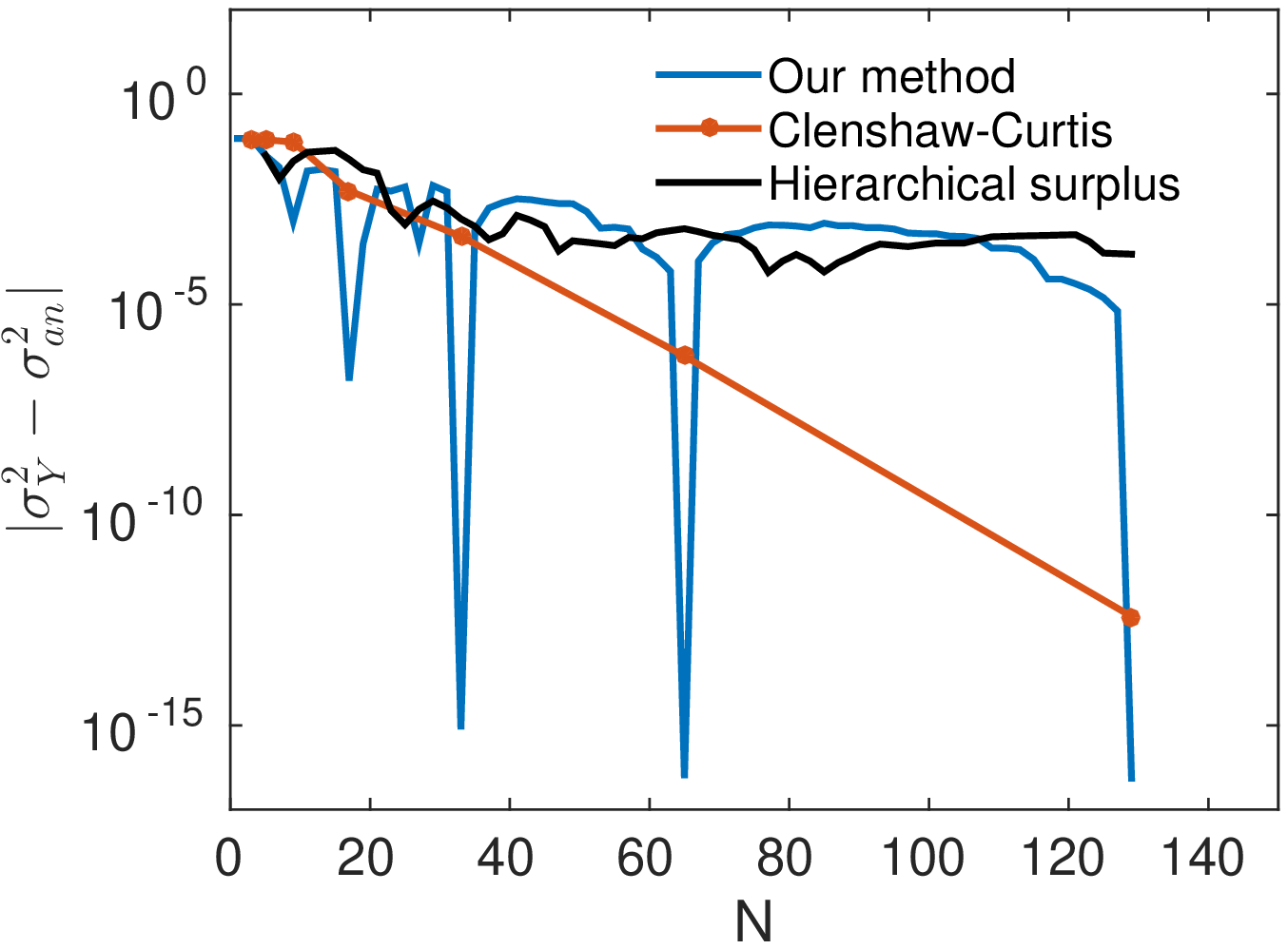}
\caption{Case 2. Absolute error in the variance $\sigma_{Y}^2$ versus number of sampling 
points $N$.  Blue: new adaptive selection method. Red: Clenshaw-Curtis. Black: adaptive method based on hierarchical surplus.} \label{fig:case_2_sigma}
\end{figure}

In this case the function $g(x)$ is periodic, and it presents, in the domain 
$x\in[-1,1]$ three local minima ($y=1/9$) and three local maxima ($y=1$).
The function and the cdf $C(y)$ are shown in Figure \ref{fig:case_2_f} (top and bottom 
panel, respectively). Figure \ref{fig:case_2_err} shows the error for this case (from now on the same format of Figure \ref{fig:case_1_err} will be used). 
The first consideration is that the hierarchical surplus method is the less accurate of the three.
Second, $\varepsilon_{g}$ is essentially the same for the CC and the new method, up to $N=65$. For 
$N=129$ the CC methods achieve a much accurate solution as compared to the 
new adaptive method, whose error has a much slower convergence. However, looking at 
the error in the cdf in top panel of Figure \ref{fig:case_2_err}, the two methods are 
essentially equivalent. This example demonstrates that, in an UQ 
framework, the primary goal in constructing a good interpolant should not be to 
minimize the error of the interpolant with respect to the 'true' $g(x)$, but 
rather to achieve the fastest possible convergence on the cdf $C_{Y}$. Although, 
the two effects are intuitively correlated, they are not into a linear 
relationship. In other words, not all sample points in $x$ count equally in 
minimizing $\varepsilon_{C}$.
The convergence of $\mu_{Y}$ (exact value $\mu_{an} = 0.385$) and $\sigma_{Y}^2$ (exact value $\sigma_{an}=0.087$) is shown 
in Figures \ref{fig:case_2_mu} 
and \ref{fig:case_2_sigma}, respectively. It is interesting to notice that our method presents errors that are always smaller than the CC method,
although the errors degrade considerably in the regions between two CC points, where the two adaptive methods yield comparable results.

\subsection{Case 3: Lotka-Volterra model (predator-prey)}

\begin{figure}
\includegraphics[width=10 cm]{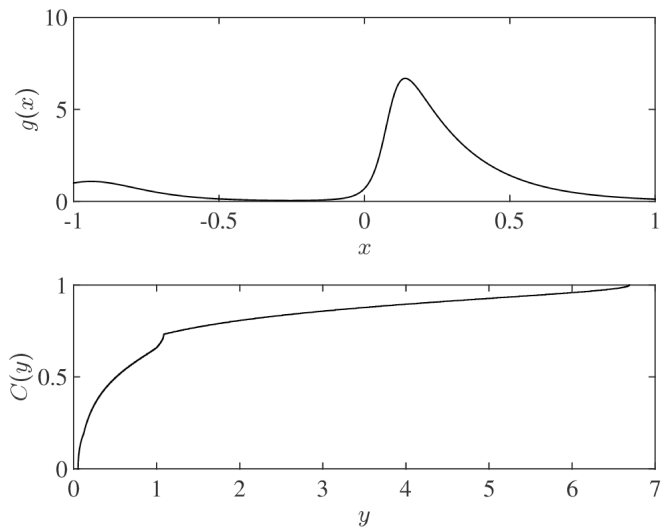}
\caption{Case 3: Lotka-Volterra model. Top panel: $g(x)$; bottom panel: 
$C(y)$.}\label{fig:case_3_f}
\end{figure}

\begin{figure}
\includegraphics[width=10 cm]{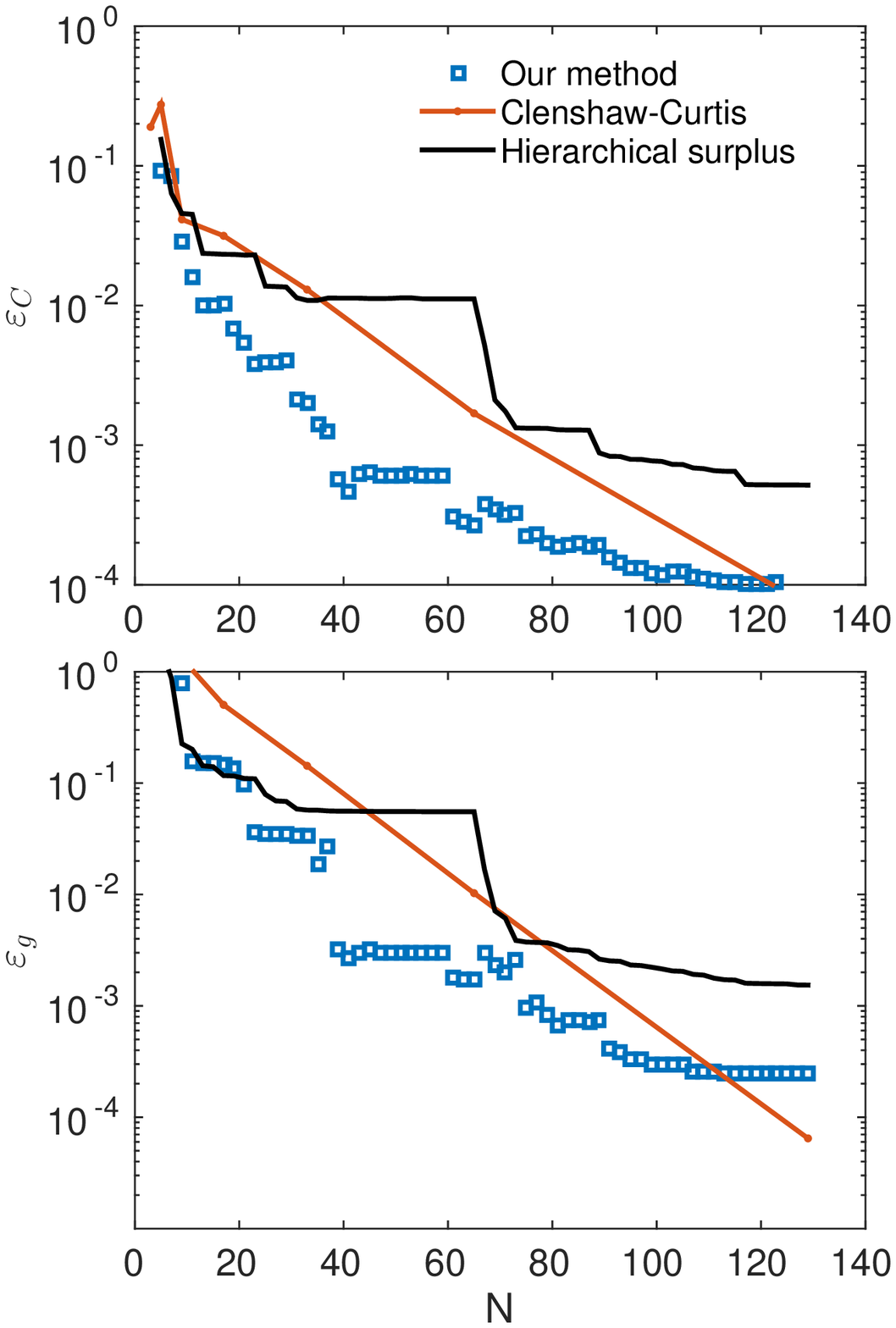}
\caption{Case 3. Error $\varepsilon_{C}$ (top) $\varepsilon_{g}$ (bottom) as function of number of sampling points 
$N$. Blue squares: new adaptive selection method. Red dots: Clenshaw-Curtis. Black curve: adaptive method based on hierarchical surplus.} \label{fig:case_3_err}
\end{figure}

\begin{figure}
\includegraphics[width=10 cm]{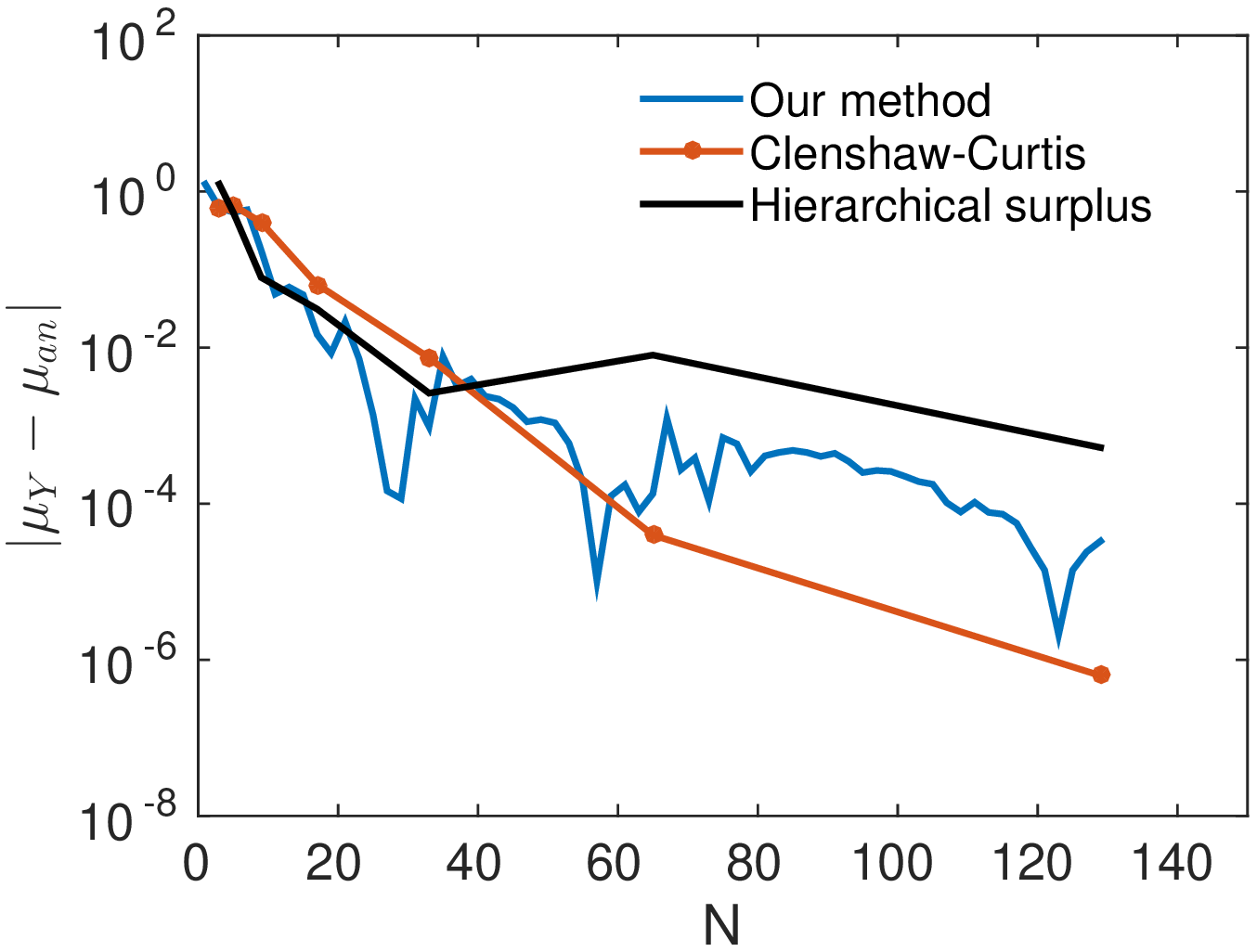}
\caption{Case 3. Absolute error in the mean $\mu_Y$ versus number of sampling 
points $N$.  Blue: new adaptive selection method. Red: Clenshaw-Curtis. Black: adaptive method based on hierarchical surplus.}\label{fig:case_3_mu}
\end{figure}

\begin{figure}
\includegraphics[width=10 cm]{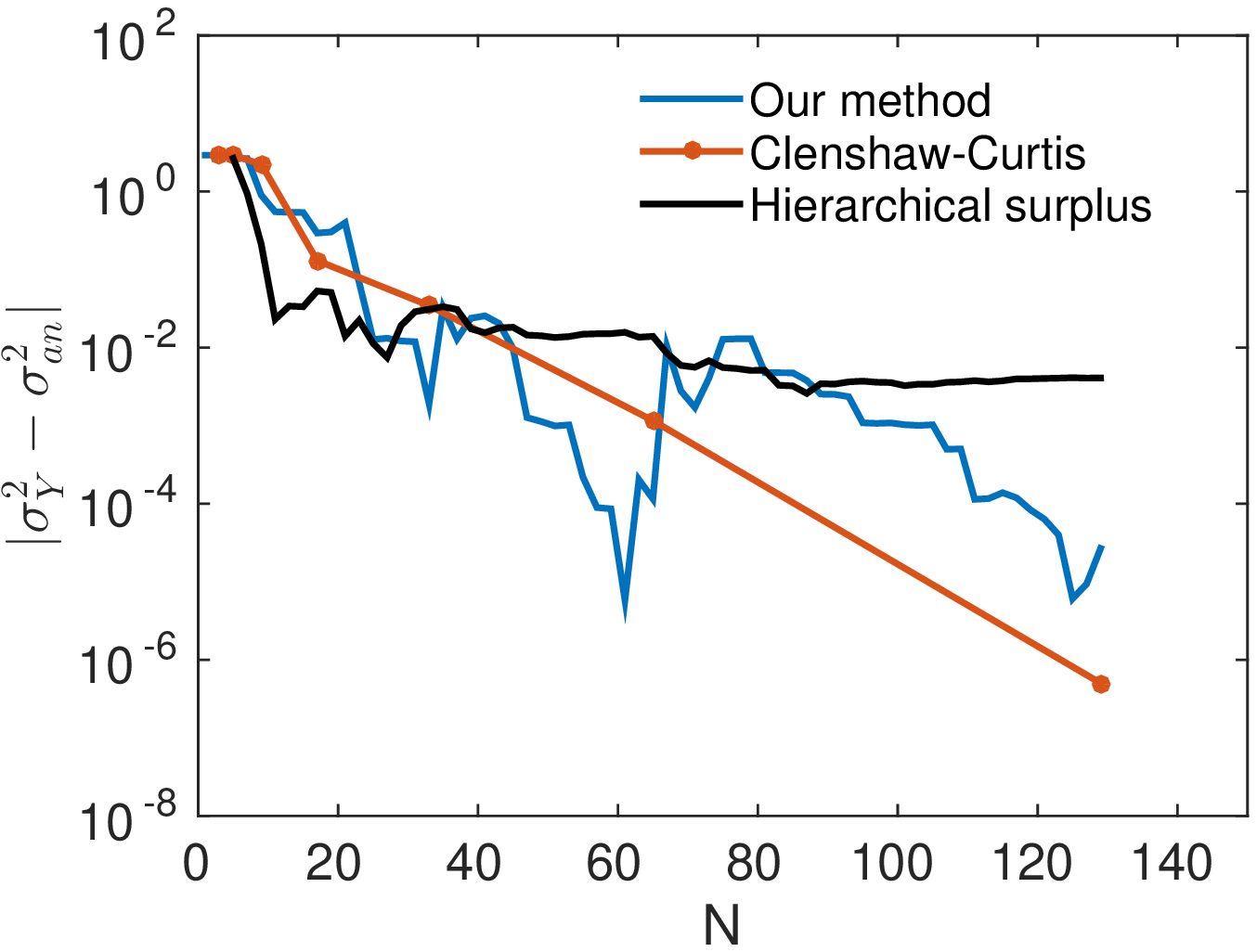}
\caption{Case 3. Absolute error in the variance $\sigma_{Y}^2$ versus number of sampling 
points $N$.  Blue: new adaptive selection method. Red: Clenshaw-Curtis. Black: adaptive method based on hierarchical surplus.}\label{fig:case_3_sigma}
\end{figure}

The Lotka-Volterra model \cite{link16,stevens2009lotka,wangersky1978lotka} is a well-studied model that exemplifies 
the interaction between two populations (predators and preys).
This case is more realistic than Cases 1 and 2, as the solution of the model 
cannot be written in analytical form. As such, both the $g(x)$ and the cdf $C(y)$ 
used to compute the errors are calculated numerically. We use the following 
simple model:
\begin{eqnarray}
 \frac{{\rm d}h(t)}{{\rm d}t} &=& h(t) - (5x+6)h(t)l(t)\\
 \frac{{\rm d}l(t)}{{\rm d}t} &=& h(t)l(t) - l(t)
\end{eqnarray}
where $h(t)$ and $l(t)$ denote the population size for each species (say, 
horses and lions) as function of time. The ODE is easily solved in MATLAB, 
with the \texttt{ode45} routine, with an absolute tolerance set equal to 
$10^{-8}$. We use, as initial conditions, $h(t=0)=l(t=0)=1$, and we solve the 
equations for $t\in[0,10]$. Clearly, the solution of the model depends on the 
input parameter $x$. We define our test function $g(x)$ to be the result of the 
model for the $l$ population at time $t=10$:
\begin{equation}
 g(x)=l(t=10,x).                                                                 
\end{equation}
The resulting function $g(x)$, and the computed cdf $C(y)$ are shown in Figure 
\ref{fig:case_3_f} (top and bottom panel, respectively). We note that, although 
$g(x)$ cannot be expressed as an analytical function, it is still smooth, and 
hence it does not present particular difficulties in being approximated through 
a polynomial interpolant. Indeed the error $\varepsilon_{g}$ undergoes a fast 
convergence both for the adaptive methods and for the CC method (Figure 
\ref{fig:case_3_err}). 
Once again, the new adaptive method is much more powerful than the CC method in 
achieving a better convergence rate, and thus saving computational power,
while the hierarchical surplus method is the worst of the three.
Convergence of $\mu_{Y}$ and $\sigma_{Y}^2$ are shown in 
Figures \ref{fig:case_3_mu} and 
\ref{fig:case_3_sigma}, respectively. Similar to previous cases, 
the CC presents a monotonic convergence, while this is not the case for
the adaptive methods. Only for $N=129$, the CC method yields much better results 
than the new method.

\subsection{Case 4: Van der Pol oscillator}

\begin{figure}
\includegraphics[width=10 cm]{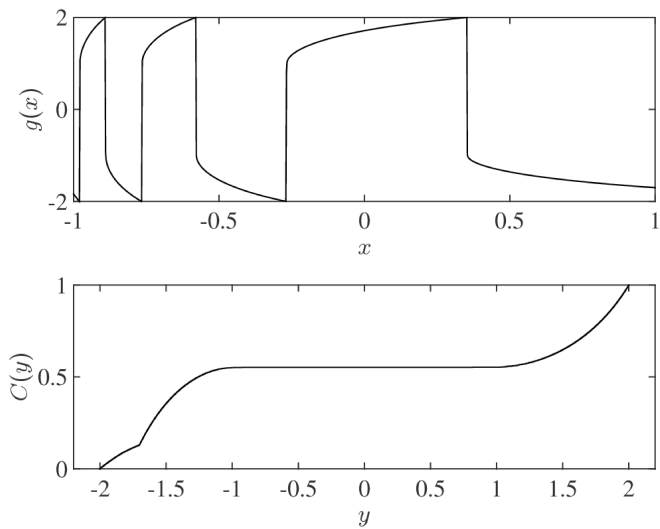}
\caption{Case 4: Van der Pol oscillator. Top panel: $g(x)$; bottom panel: 
$C(y)$. }\label{fig:case_4_f}
\end{figure}

\begin{figure}
\includegraphics[width=10 cm]{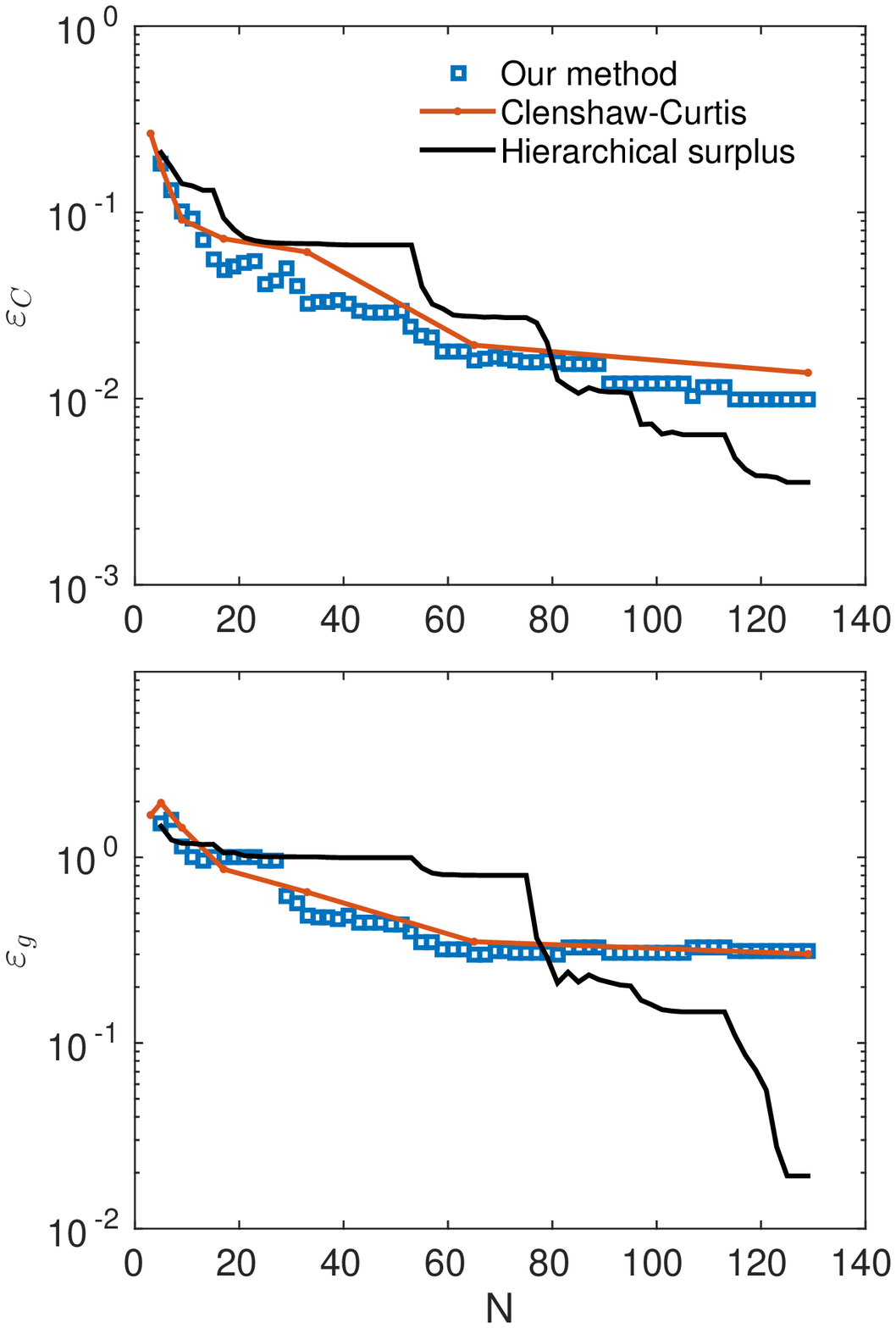}
\caption{Case 4. Error $\varepsilon_{C}$ (top) $\varepsilon_{g}$ (bottom) as function of number of sampling points 
$N$. Blue squares: new adaptive selection method. Red dots: Clenshaw-Curtis. Black curve: adaptive method based on hierarchical surplus.}\label{fig:case_4_err}
\end{figure}

\begin{figure}
\includegraphics[width=10 cm]{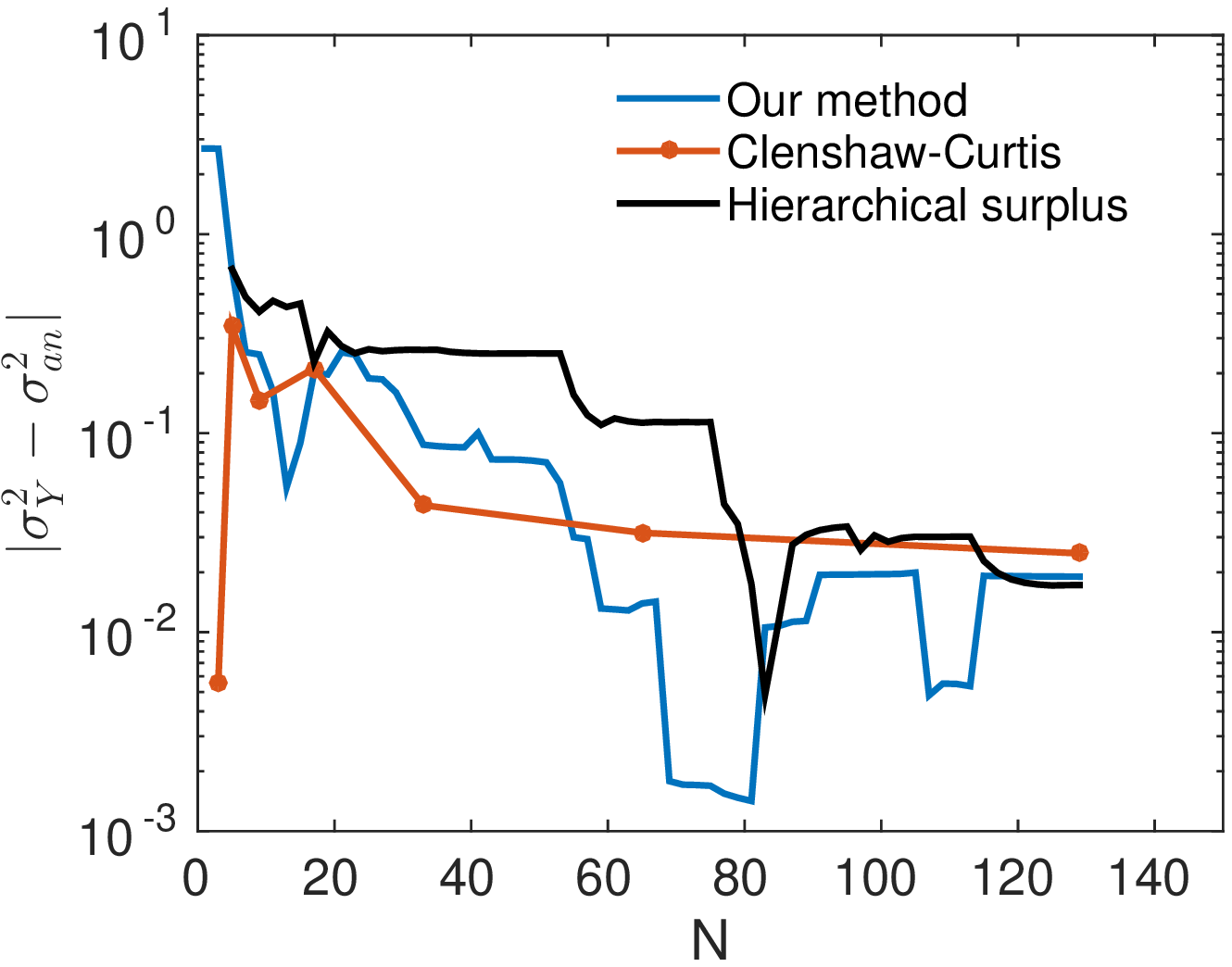}
\caption{Case 4. Absolute error in the variance $\sigma_{Y}^2$ versus number of sampling 
points $N$.  Blue: new adaptive selection method. Red: Clenshaw-Curtis. Black: adaptive method based on hierarchical surplus.}\label{fig:case_4_sigma}
\end{figure}

Our last example is the celebrated Van der Pol oscillator\cite{link15,eldred2009recent,PhysRevE.87.062109,liu2007novel}, which 
has been extensively studied as a textbook case of a nonlinear dynamical 
system. 
In this respect this test case is very relevant to Uncertainty Quantification, 
since real systems often exhibit a high degree of nonlinearity.
Similar to Case 3, we define our test function $g(x)$ as the output of a set 
of two ODEs, which we solve numerically with MATLAB.
The model for the Van der Pol oscillator is:
\begin{eqnarray}
\frac{{\rm d}Q(t)}{{\rm d}t} &=& V(t)\\
\frac{{\rm d}V(t)}{{\rm d}t} &=& (-50+100(x+2))(1 - Q(t)^2)V(t) -Q(t).
\end{eqnarray}
The initial conditions are $Q(t=0)= 2$, $V(t=0)=0$. The model is solved for time 
$t\in[0,300]$, and the function $g(x)$ is defined as 
\begin{equation}
 g(x)=V(t=300,x).                                                                
\end{equation}
The so-called nonlinear damping parameter is rescaled such that for 
$x\in[-1,1]$, it ranges between 50 and 250. The function $g(x)$ and the 
corresponding cdf $C(y)$ are shown in Figure \ref{fig:case_4_f}. This function is 
clearly much more challenging than the previous ones. It is divided in two 
branches, where it takes values $-2\leq y\leq-1$ and $1\leq y \leq2$, and it 
presents discontinuities where it jumps from one branch to the other. 
Correspondingly, cdf $C(y)$ presents a flat plateau for $-1\leq 
y\leq 1$, which is the major challenge for both methods.
In figure \ref{fig:case_4_err} we show the errors 
$\varepsilon_{g}$ and $\varepsilon_{C}$. The overall 
convergence rate of the CC and the new method is similar.
For this case, the hierarchical surplus method yields a better convergence, 
but only for $N>80$.
As we commented before, the mean $\mu_{Y}$ has no statistical 
meaning in this  case, because the output is divided into two separate regions. The convergence 
for $\sigma_{Y}^2$ is presented in Figure \ref{fig:case_4_sigma}.

\section{Conclusions and future work}
We have presented a new adaptive algorithm for the selection of sampling 
points for non-intrusive stochastic collocation in Uncertainty Quantification (UQ). 
The main idea is to use a radial basis function as interpolant, and to refine the grid on points where the interpolant presents 
large and small first derivative.\\
In this work we have focused on 1D and uniform probability $P_X(x)$, and
 we have shown four test cases, encompassing analytical and non-analytical 
smooth functions, which are prototype of a very wide class of functions. In all 
cases the new adaptive method improved the efficiency of both the (non-adaptive) Clenshaw-Curtis collocation method,
and of the adaptive algorithm based on the calculation of the hierarchical surplus (note that the method used in this paper is a
slight improvement of the classical algorithm).
The strength of our method is the 
ability to select a new sampling point making full use of the interpolant 
resulting from all the previous evaluation of the function $g(x)$, thus seeking 
the most optimal convergence rate for the cdf $C(y)$. We 
have shown that there is no one-to-one correspondence between a reduction in the 
interpolation error $\varepsilon_{g}$ and a reduction in the cdf error 
$\varepsilon_{C}$. For this reason, collocation methods that choose the 
distribution of sampling points a priori can perform poorly in attaining a fast 
convergence rate in $\varepsilon_{C}$, which is the main goal of UQ. 
Moreover, in order to maintain the nestedness of the collocation 
points the CC method requires larger and larger number of simulations ($2^w$ 
moving from level $w$ to level $w+1$), which is in contrast with our new method 
where one can add one point at the time.\\
We envision many possible research directions to further investigate our method.
The most obvious is to study multi-dimensional problems. We emphasize that the 
radial basis function is a mesh-free method and as such we anticipate that this 
will largely alleviate the curse of dimensionality that afflicts other 
collocation methods based on quadrature points (however, see \cite{link17} for 
methods related to the construction of sparse grids, which have the same aim).
Moreover, it will be interesting to explore the versatility of RBF in what 
concerns the possibility of choosing an optimal shape parameter $c$ 
\cite{link18}. Recent work \cite{jung2009iterative,wang2002optimal} investigated the role of the shape parameter $c$ 
in interpolating discontinuous functions, which might be very relevant in the 
context of UQ, when the continuity of $g(x)$ cannot be 
assumed a priori. Finally, a very appealing research direction, would be to 
simultaneously exploit quasi-Monte Carlo and adaptive selection methods for 
extremely large dimension problems.


\acknowledgements
A. A. and C. R. are supported by FOM Project No. 67595 and 12PR304, respectively. We would like to acknowledge dr.ir. J.A.S. Witteveen ($\dagger$ 2015) for the useful discussions we had about uncertainty quantification.











\bibliographystyle{IJ4UQ_Bibliography_Style}


\begin{thebibliography}{10}

\bibitem{xiu09}
Xiu, D., Fast numerical methods for stochastic computations: a review, {\em
  Communications in computational physics}, 5(2-4):242--272, 2009.

\bibitem{eldred09}
Eldred, M. and Burkardt, J., Comparison of non-intrusive polynomial chaos and
  stochastic collocation methods for uncertainty quantification, {\em AIAA
  paper}, 976(2009):1--20, 2009.

\bibitem{onorato2010comparison}
Onorato, G., Loeven, G., Ghorbaniasl, G., Bijl, H., and Lacor, C., Comparison
  of intrusive and non-intrusive polynomial chaos methods for cfd applications
  in aeronautics, In {\em V European Conference on Computational Fluid Dynamics
  ECCOMAS, Lisbon, Portugal}, pp. 14--17, 2010.

\bibitem{xiu02}
Xiu, D. and Karniadakis, G.E., The wiener--askey polynomial chaos for
  stochastic differential equations, {\em SIAM journal on scientific
  computing}, 24(2):619--644, 2002.

\bibitem{crestaux09}
Crestaux, T., Le~Ma{\i}ˆtre, O., and Martinez, J.M., Polynomial chaos
  expansion for sensitivity analysis, {\em Reliability Engineering \& System
  Safety}, 94(7):1161--1172, 2009.

\bibitem{Togawa:2011}
Togawa, K., Benigni, A., and Monti, A., Advantages and challenges of
  non-intrusive polynomial chaos theory, In {\em Proceedings of the 2011 Grand
  Challenges on Modeling and Simulation Conference}, GCMS '11, pp. 30--35,
  Vista, CA, 2011. Society for Modeling \& Simulation International.

\bibitem{eldred2009recent}
Eldred, M.S., Recent advances in non-intrusive polynomial chaos and stochastic
  collocation methods for uncertainty analysis and design, {\em AIAA Paper},
  2274:2009, 2009.

\bibitem{link2}
Tempone, I.B.I.R. and G.Zouraris, Galerkin finite element approximations of
  stochastic elliptic partial differential equations, {\em SIAM Journal of
  Numerical Analysis}, 42(2):800--825, 2004.

\bibitem{grigoriu2012stochastic}
Grigoriu, M., {\em Stochastic systems: uncertainty quantification and
  propagation}, Springer Science \& Business Media, 2012.

\bibitem{xiu2010numerical}
Xiu, D., {\em Numerical methods for stochastic computations: a spectral method
  approach}, Princeton University Press, 2010.

\bibitem{le10}
Le~Ma{\^\i}tre, O. and Knio, O.M., {\em Spectral methods for uncertainty
  quantification: with applications to computational fluid dynamics}, Springer
  Science \& Business Media, 2010.

\bibitem{link4}
Caflisch, R.E., Monte carlo and quasi-monte carlo methods, {\em Acta Numerica},
  7:1--49, 1998.

\bibitem{kalos2008monte}
Kalos, M.H. and Whitlock, P.A., {\em Monte carlo methods}, John Wiley \& Sons,
  2008.

\bibitem{l2009monte}
L'Ecuyer, P. and Owen, A.B., {\em Monte Carlo and Quasi-Monte Carlo Methods
  2008}, Springer, 2009.

\bibitem{link5}
Niederreiter, H.G., Quasi-monte carlo methods and pseudo-random numbers, {\em
  Bull. Amer. Math. Soc.}, 84(6):957--1041, 1978.

\bibitem{kalos2008pseudorandom}
Kalos, M.H. and Whitlock, P.A., Pseudorandom numbers, {\em Monte Carlo Methods,
  Second Edition}, pp. 179--197, 2008.

\bibitem{link6}
Bellman, R., Dynamic programming, {\em Courier Dover Publications}, 2003.

\bibitem{indyk1998approximate}
Indyk, P. and Motwani, R., Approximate nearest neighbors: towards removing the
  curse of dimensionality, In {\em Proceedings of the thirtieth annual ACM
  symposium on Theory of computing}, pp. 604--613. ACM, 1998.

\bibitem{kuo2005lifting}
Kuo, F.Y. and Sloan, I.H., Lifting the curse of dimensionality, {\em Notices of
  the AMS}, 52(11):1320--1328, 2005.

\bibitem{link7}
Xiu, D. and Hesthaven, J.S., High-order collocation methods for differential
  equations with random inputs, {\em SIAM J. Sci. Comput.}, 28:1167--1185,
  2006.

\bibitem{Foo2010}
Foo, J. and Karniadakis, G.E., Multi-element probabilistic collocation method
  in high dimensions, {\em J. Comput. Phys.}, 229(5):1536--1557, March 2010.

\bibitem{babuvska2007stochastic}
Babu{\v{s}}ka, I., Nobile, F., and Tempone, R., A stochastic collocation method
  for elliptic partial differential equations with random input data, {\em SIAM
  Journal on Numerical Analysis}, 45(3):1005--1034, 2007.

\bibitem{wang12}
Wang, P. and Tartakovsky, D.M., Uncertainty quantification in kinematic-wave
  models, {\em Journal of computational Physics}, 231(23):7868--7880, 2012.

\bibitem{wang13}
Wang, P., Tartakovsky, D.M., Jarman~Jr, K., and Tartakovsky, A.M., Cdf
  solutions of buckley--leverett equation with uncertain parameters, {\em
  Multiscale Modeling \& Simulation}, 11(1):118--133, 2013.

\bibitem{smolyak63}
Smolyak, S.A., Quadrature and interpolation formulas for tensor products of
  certain classes of functions, In {\em Dokl. Akad. Nauk SSSR}, Vol.~4, p. 123,
  1963.

\bibitem{ganapathysubramanian07}
Ganapathysubramanian, B. and Zabaras, N., Sparse grid collocation schemes for
  stochastic natural convection problems, {\em Journal of Computational
  Physics}, 225(1):652--685, 2007.

\bibitem{bieri09}
Bieri, M., Andreev, R., and Schwab, C., Sparse tensor discretization of
  elliptic spdes, {\em SIAM Journal on Scientific Computing}, 31(6):4281--4304,
  2009.

\bibitem{ma09b}
Ma, X. and Zabaras, N., An efficient bayesian inference approach to inverse
  problems based on an adaptive sparse grid collocation method, {\em Inverse
  Problems}, 25(3):035013, 2009.

\bibitem{link17}
Tempone, F.N.R. and Webster, C.G., A sparse grid stochastic collocation method
  for partial differential equations with random input data, {\em SIAM J.
  Numer. Anal.}, 46(5):2309--2345, 2008.

\bibitem{jakeman12}
Jakeman, J.D. and Roberts, S.G.
\newblock Local and dimension adaptive stochastic collocation for uncertainty
  quantification.
\newblock In {\em Sparse grids and applications}, pp. 181--203. Springer, 2012.

\bibitem{nguyen12}
Nguyen, N.H., Willcox, K., and Khoo, B.C., Model order reduction for stochastic
  optimal control, In {\em ASME 2012 11th Biennial Conference on Engineering
  Systems Design and Analysis}, pp. 599--606. American Society of Mechanical
  Engineers, 2012.

\bibitem{link8}
Laurie, P.D., Computation of gauss-type quadrature formulas, {\em J. Comp.
  Appl. Math.}, 127(1-2):201--217, 2001.

\bibitem{gautschi1970construction}
Gautschi, W., On the construction of gaussian quadrature rules from modified
  moments., {\em Mathematics of Computation}, 24(110):245--260, 1970.

\bibitem{waldvogel2006fast}
Waldvogel, J., Fast construction of the fej{\'e}r and clenshaw--curtis
  quadrature rules, {\em BIT Numerical Mathematics}, 46(1):195--202, 2006.

\bibitem{link9}
Berrut, J.P. and Trefethen, L.N., Barycentric lagrange interpolation, {\em SIAM
  Rev.}, 46(3):501--517, 2004.

\bibitem{link19}
Trefethen, L.N., Spectral methods in matlab, {\em SIAM: Society for Industrial
  and Applied Mathematics}, 2000.

\bibitem{link11}
Epperson, J., On the runge example, {\em Amer. Math. Monthly}, 94:329--341,
  1987.

\bibitem{ma09}
Ma, X. and Zabaras, N., An adaptive hierarchical sparse grid collocation
  algorithm for the solution of stochastic differential equations, {\em Journal
  of Computational Physics}, 228(8):3084--3113, 2009.

\bibitem{witteveen12}
Witteveen, J.A. and Iaccarino, G., Refinement criteria for simplex stochastic
  collocation with local extremum diminishing robustness, {\em SIAM Journal on
  Scientific Computing}, 34(3):A1522--A1543, 2012.

\bibitem{link10}
Boyd, J.P., Chebychev and fourier spectral methods, 2nd ed., {\em Dover, New
  York}, 2001.

\bibitem{link16}
Baruer, F. and Castillo-Chavez, C., Mathematical models in population biology
  and epidemiology, {\em Springer-Verlag}, 2000.

\bibitem{stevens2009lotka}
Stevens, M.H.H.
\newblock Lotka--volterra interspecific competition.
\newblock In {\em A Primer of Ecology with R}, pp. 135--159. Springer, 2009.

\bibitem{wangersky1978lotka}
Wangersky, P.J., Lotka-volterra population models, {\em Annual Review of
  Ecology and Systematics}, pp. 189--218, 1978.

\bibitem{link15}
Geer, M.B.D.J. and Andersen, C.M., Perturbation analysis of the limit cycle of
  the free van der pol equation, {\em SIAM Journal on Applied Mathematics},
  44(5):881--895, 1984.

\bibitem{PhysRevE.87.062109}
Yuan, R., Wang, X., Ma, Y., Yuan, B., and Ao, P., Exploring a noisy van der pol
  type oscillator with a stochastic approach, {\em Phys. Rev. E}, 87:062109,
  Jun 2013.

\bibitem{liu2007novel}
Liu, L., Dowell, E.H., and Hall, K.C., A novel harmonic balance analysis for
  the van der pol oscillator, {\em International Journal of Non-Linear
  Mechanics}, 42(1):2--12, 2007.

\bibitem{link18}
Fasshauer, G.E. and Zhang, J.G., On choosing ``optimal'' shape parameters for
  rbf approximation, {\em Numer. Algor.}, 45:345--368, 2007.

\bibitem{jung2009iterative}
Jung, J.H. and Durante, V.R., An iterative adaptive multiquadric radial basis
  function method for the detection of local jump discontinuities, {\em Applied
  Numerical Mathematics}, 59(7):1449--1466, 2009.

\bibitem{wang2002optimal}
Wang, J. and Liu, G., On the optimal shape parameters of radial basis functions
  used for 2-d meshless methods, {\em Computer methods in applied mechanics and
  engineering}, 191(23):2611--2630, 2002.

\end{thebibliography}

\newpage

\begin{figure}
\includegraphics[width=10 cm]{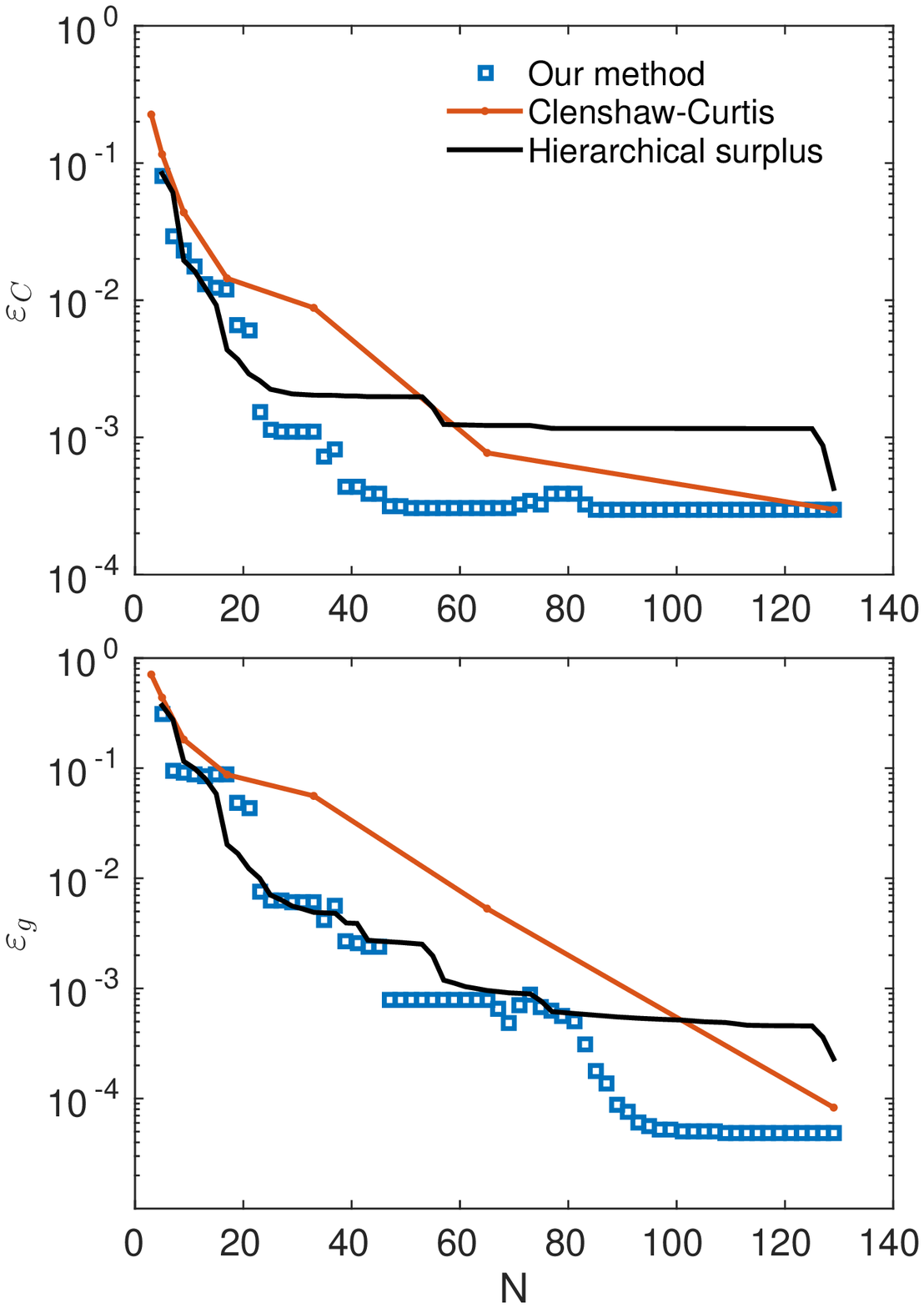}
\caption{Case 1. Error $\varepsilon_{C}$ (top) $\varepsilon_{g}$ (bottom) as function of number of sampling points 
$N$. Blue squares: new adaptive selection method. Red dots: Clenshaw-Curtis. Black curve: adaptive method based on hierarchical surplus.} \label{fig:case_1_err}
\end{figure}

\end{document}